\documentclass[preprint,12pt,a4paper]{elsarticle}
\usepackage[utf8]{inputenc}
\usepackage{mathtools}
\usepackage{amsmath, amssymb, amsthm, bm}  
\usepackage[frozencache,cachedir=.]{minted}
\usepackage{placeins}

\usepackage[smaller]{acronym}
\usepackage{svg}
\usepackage{url}

\usepackage[noabbrev,capitalize]{cleveref}

\crefname{equation}{}{}
\Crefname{equation}{}{}

\theoremstyle{definition} 

\newtheorem{remark}{Feature Remark}

\theoremstyle{plain} 
\theoremstyle{remark} 

\definecolor{bg}{rgb}{0.95,0.95,0.95}

\begin{document}
\sloppy 

\title{Power Market Tool (POMATO)\\ for the Analysis of Zonal Electricity Markets}


\author[TUWIP]{Richard Weinhold\corref{mycorrespondingauthor}\fnref{eqconfn}}
\ead{riw@wip.tu-berlin.de}

\author[NYUTandon,TUContr]{Robert Mieth\fnref{eqconfn}}
\ead{robert.mieth@nyu.edu}

\cortext[mycorrespondingauthor]{Corresponding author}

\fntext[eqconfn]{R. Weinhold and R. Mieth contributed equally to this work.}

\address[TUWIP]{Fakultät VII Wirtschaft und Management, TU Berlin, 10623 Berlin, Germany.}
\address[NYUTandon]{Department of Electrical and Computer Engineering, Tandon School of Engineering,\\ New York University, New York, NY 10012 USA.}
\address[TUContr]{Fakultät IV Elektrotechnik und Informatik, TU Berlin, 10587 Berlin, Germany.}

\begin{frontmatter}
\begin{abstract}
The proposed open-source \emph{Power Market Tool} ({\small{POMATO}}) aims to enable research on interconnected modern and future electricity markets in the context of the physical transmission system and its secure operation. POMATO has been designed to study capacity allocation and congestion management (CACM) policies of European zonal electricity markets, especially flow-based market coupling (FBMC). 
For this purpose, {\small{POMATO}} implements methods for the analysis of simultaneous zonal market clearing, nodal (N-k secure) power flow computation for capacity allocation, and multi-stage market clearing with adaptive grid representation and redispatch.  
The computationally demanding N-k secure power flow is enabled via an efficient constraint reduction algorithm. 
POMATO provides an integrated environment for data read-in, pre- and post-processing and interactive result visualization. 
Comprehensive data sets of European electricity systems compiled from Open Power System Data and Matpower Cases are part of the distribution. 
{\small{POMATO}} is implemented in Python and Julia, leveraging Python's easily maintainable data processing and user interaction features and Julia's well readable algebraic modeling language, superior computational performance and interfaces to open-source and commercial solvers.
\end{abstract}


\end{frontmatter}

\begin{table}[H]
\resizebox{\textwidth}{!}{
\begin{tabular}{|l|p{6.5cm}|p{6.5cm}|}
\hline
\textbf{Nr.} & \textbf{Code metadata description} & \textbf{Please fill in this column} \\ \hline
C1 & Current code version & 0.3 \\
\hline
C2 & Permanent link to code/repository used for this code version & \url{https://github.com/richard-weinhold/pomato} \\ \hline
C3 & Code Ocean compute capsule & \\ \hline
C4 & Legal Code License   & LGPL v3 \\ \hline
C5 & Code versioning system used & git \\ \hline
C6 & Software code languages, tools, and services used & Python and Julia \\ \hline
C7 & Compilation requirements, operating environments \& dependencies &  \\ \hline
C8 & If available Link to developer documentation/manual & \url{https://pomato.readthedocs.io/en/latest/} \\ \hline
C9 & Support email for questions & \url{riw@wip.tu-berlin.de} \\ \hline
\end{tabular}
}
\caption{Code metadata (mandatory)}
\label{tbl:code_metadata} 
\end{table}

\section{Motivation, Significance and Impact}

Europe's increasing electricity production from renewable energy resources in combination with a significant decline of conventional generation capacity, has spawned political and academic interest in the transmission system's ability to accommodate this transition, \cite{amprion_fbmc}.
Central to this discussion is the efficiency of \textit{capacity allocation} and \textit{congestion management} (CACM) policies between and within electricity market areas that share transmission infrastructure, \cite{eu_2015_1222}.  
Capacity allocation (CA) summarizes regulatory and market mechanisms that constrain electricity trading volumes between two adjacent market areas with respect to the expected available cross-border transmission capacity at time of delivery. 
Congestion management (CM), on the other hand, refers to methods ensuring that the physical system state at time of delivery indeed remains within its security margins, e.g. does not cause transmission line overloads. 
Noticeably, well defined CA based on suitable forecasts can reduce CM measures, such as out-of-market generator redispatch coordinated by the transmission system operator (TSO).
Previously implemented CA policies based on static \emph{net-transfer capacities} (NTCs) consider tie-line capacities between markets, but neglect restricting transmission assets within market zones, thus leading to often non-feasible market outcomes, increased CM (redispatch) or overly conservative market results, \cite{amprion2011,ec2016/1719}.
To overcome these deficiencies and in an effort to "\textit{move towards a genuinely integrated [European] electricity market}", \cite{eu_2015_1222}, the central-western European (CWE) countries inaugurated \emph{flow-based market coupling} (FBMC), a more complex CA policy that aims to increase the potential volume of cross-border electricity trading while decreasing CM requirements by explicitly accounting for cross-border and zone-internal transmission limits.

The proposed \emph{Power Market Tool} ({\small{POMATO}}) has been designed to enable further research on the status-quo and future policies of practical zonal electricity markets, especially FBMC.
While the theory of a centrally coordinated zonal electricity market is somewhat mature, see e.g. \cite{ehrenmann2005inefficiencies}, practical implementation with imperfect coordination between market and system operators requires ongoing analyses.
The current European FBMC is a multi-stage process coordinated by multiple TSOs and involves detailed zone-specific load and generation forecasts and network models, which are typically not, or only partially, disclosed by the TSOs, \cite{schonheit2020impact}.
However, the medium- and long-term evolution of the FBMC design requires an informed public decision based on independent FBMC analyses that study the impact of, e.g., more countries joining the coupled market or bidding zone layouts (which have been declared inefficient by a recent TSO study, \cite{entsoe_biddingzones}). 
Notably, the yearly federal report on the future of the grid in Germany (\emph{``Netzentwicklungsplan''}), included a rudimentary FBMC representation for the first time in its 2018 edition, \cite{fur2018netzentwicklungsplanung}, three years after FBMC implementation.

There are few model implementations of the FBMC process available to the academic community, \cite{aravena2016renewable,matthes2019impact,schonheit2020elmod}.
In \cite{aravena2016renewable} Aravena et al. compare various CA policies and demonstrate the benefits of FBMC traditional approaches using real-world data sets.
Also, this study proposes an iterative approach to meet the practical requirement that any dispatch has to be robust against unplanned transmission equipment outage (\emph{N-1 security}). However, data-sets and model implementation are not published along the paper. 
Similarly, Matthes et al. \cite{matthes2019impact} present a FBMC formulation for estimating future FBMC parameters and studying the impact of regulations that require additional security margins on critical transmission lines. This model uses an external convex hull reduction for a N-1 secure FBMC solution and is implemented in the MILES framework, \cite{spieker2016european}, which, to our knowledge, is also not publicly available. 
Schönheit et al. \cite{schonheit2020elmod} extend the classic  GAMS\footnote{~\url{www.gams.com}} model ELMOD \cite{leuthold2012large} to facilitate FBMC analyses and study the impact of regulations requiring a minimum availability of interzonal trading volumes, \cite{schonheit2020minimum}.
Here, N-1 security is approximated through static security margins and despite comprehensive documentation in \cite{schonheit2020elmod} the code extension itself is undisclosed. 

{\small{POMATO}} aims to overcome some of the caveats of \cite{aravena2016renewable,matthes2019impact,schonheit2020elmod} by providing a documented and open-source framework that is easy to use and goes beyond the implementation of an academic optimization model. 
Some main features and contributions are:
\begin{enumerate}[(a)]
     \item Separation of data processing (implemented in Python\footnote{~\url{www.python.org}}) and optimization (implemented in Julia\footnote{~\url{www.julialang.org}}) to achieve a flexible Python-based user interface that is familiar to many users, while creating a lean implementation of the central optimization model in the well-readable JuMP algebraic modeling language, \cite{DunningHuchetteLubin2017}. 
    
    \item Open-source available and documented on GitHub\footnote{~\url{github.com/richard-weinhold/pomato}}, \cite{pomato_repo,pomato_doc}.
    
    \item Comprehensive data set of the European electricity market and transmission infrastructure based on Open Power Systems Data\footnote{~\url{open-power-system-data.org/}} and \mbox{Matpower}\footnote{~\url{matpower.org/}} data sets.
    
    \item Electricity market model with zonal and nodal market clearing and a module to synthesize the FBMC process, including heat sector coupling. The underlying mathematical model has been reported in our previous work \cite{schonheit2020impact}.
    
    \item Exact N-1 secure dispatch implementation suitable for large-scale networks and multi-period analyses. The used algorithm removes redundant constraints, similar to the convex hull approach in \cite{matthes2019impact}, but has been optimized for optimal power flow (OPF) analyses, yielding improved computation times. This algorithm has been presented in our previous work \cite{weinhold2020fast}. 
    
    \item Stochastic OPF using chance-constraints to analyse the impact of forecast errors from renewable energy sources. 
\end{enumerate}

\section{Software Description}

\begin{figure}[H]
	\centering
    \includegraphics[width=0.95\textwidth]{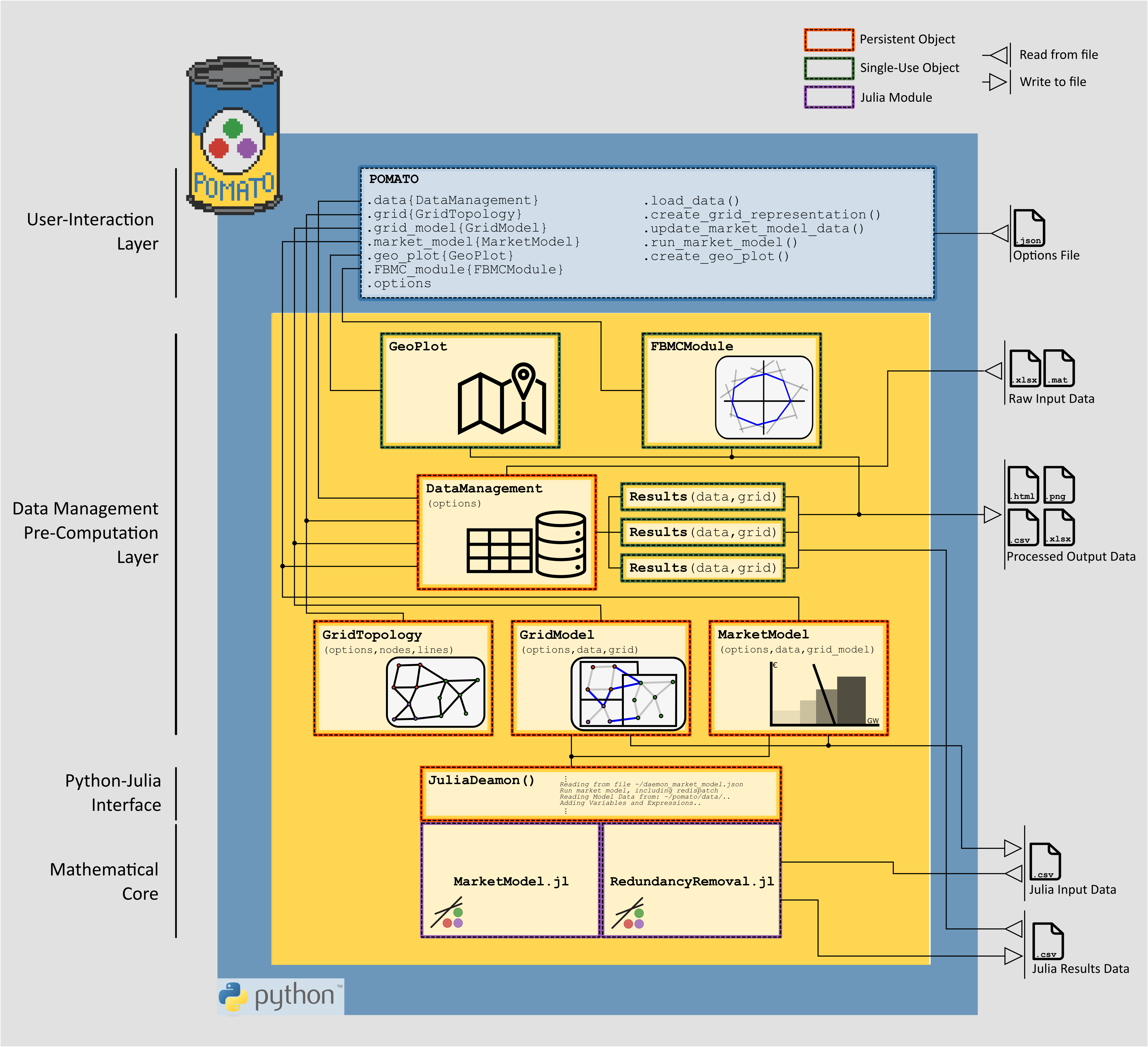}
	\caption{{\small{POMATO}}'s layered architecture. The inner model core is colored purple, the data processing layer is colored yellow, and the outer user interface is colored blue.}
	\label{fig:flowchart}
\end{figure}

{\small{POMATO}}'s architecture is structured in three layers as shown in Figure~\ref{fig:flowchart}. 
The \emph{model core} collects the mathematical formulations of the necessary optimization problems and provides an interface to the required solvers.   
To ensure a lean model implementation and efficient re-runs without re-calculating large parameter sets, the model core is encapsulated in a \emph{data processing} layer. 
This layer automates parameter calculation and validation, provides parameters to the model core, and validates and processes the resulting model output. 
Finally, all functionality of {\small{POMATO}} is collected in the outer \emph{user interface} layer via readable API-like commands.

\subsection{Model Core}\label{ssec:model_core}

The core of {\small{POMATO}} is an electricity market model.
This model is an optimization problem that finds the least cost generation schedule that matches system demand with respect to all technical constraints.
Equation~\eqref{eq:general_dispatch} shows the general structure of such a model:
\begin{subequations}
\begin{align}
  \min \text{ OBJ} &= \sum \mathit{COST\_G} + \mathit{COST\_H} + \mathit{COST\_CURT} + \mathit{OOM\_PEN} \label{eq:obj}\\ 
  \text{s.t. }& \nonumber\\
  & \text{Cost Definition} \label{eq:cost_def} \\
  & \text{Generation Constraints} \label{eq:generation}\\ 
  & \text{Heat Constraints} \label{eq:heat} \\
  & \text{Storage Constraints} \label{eq:storage}\\
  & \text{Energy Balances} \label{eq:eb} \\
  & \text{Network Constraints}\label{eq:network}.
\end{align}
\label{eq:general_dispatch}%
\end{subequations}%
Objective \eqref{eq:obj} minimizes the cost of electricity generation ($\mathit{COST\_G}$), heat generation ($\mathit{COST\_H}$), load curtailment ($\mathit{COST\_CURT}$) and out-of market penalties ($\mathit{OOM\_PEN}$, e.g. cost of redispatch) over all units and time steps.
Constraints in \eqref{eq:cost_def} define these cost components as sets of linear functions. Depending on the modeled market framework, some cost may be zero.
Constraints \eqref{eq:generation} and \eqref{eq:heat} enforce limits on the available electricity and heat generation, respectively.
Constraints \eqref{eq:storage} enforce the time-coupled energy constraints for storages and limit their efficiency-corrected charging and discharging power.
Balancing constraints in \eqref{eq:eb} ensure that for all market areas generation equals demand plus net export (export minus import) at all times.
Network constraints \eqref{eq:network} capture the transmission system topology as nodes (buses) and lines and enforce exchange capacity limits.
Physical flows are modeled using some modification of the linearized (DC) power flow equations so that nodal net injections can be mapped to power flows via a power transfer distribution factor (PTDF) matrix or voltage angles, see e.g. \cite{weinhold2020fast, horsch2018}.

The specific definitions of constraints \eqref{eq:cost_def} -- \eqref{eq:network} are activated and parametrized by {\small{POMATO}} ``on the fly'' based on user-defined options. 
For example, nodal market clearing enforces an energy balance in \eqref{eq:eb} for each \textit{node} and exchanges are limited by the physical power flow and the thermal transmission system capacities.
Zonal markets, on the other hand, ensure an aggregated energy balance for an entire \textit{zone} (i.e. multiple connected nodes).
Exchanges with neighboring zones are limited either explicitly by NTCs, which constrain individual cross-border exchanges, or implicitly through constraints on the net positions (total net export) of each zone. 
The latter is used to model FBMC by calculating zonal PTDFs, see e.g. \cite{schonheit2020impact}, that map zonal net positions to power flows on (critical) lines. 
Additionally, constraints to model CM (redispatch) and external security requirements (contingency robustness) can be included.
For redispatch models, a nodal market is solved subject to additional out-of-market penalties based on results of a previously solved zonal market. Specifically, the optimal redispatch is the least-cost change from a given generation schedule with predefined out-of-market penalties and fixed \textit{zonal} balances so that the resulting \textit{nodal} balances are feasible for the given transmission network.
Preventive system security requirements based on contingency ($N-1$) analyses, are enforced through a suitable extensions of zonal and nodal PTDF matrices.

Optimal dispatch with preventive robustness against contingencies (unplanned outages) is computationally demanding and requires additional effort, as the number of constraints grows exponentially with the network size which often renders the problem unsolvable for multi-period economic analyses, \cite{weinhold2020fast}. 
However, it has been shown that many (in fact most) of these constraints are \textit{redundant}, i.e. never binding in the optimal solution due to the existence of more restrictive constraints, \cite{bouffard2005umbrella}.
To ensure feasible solution times for real-world networks over non-trivial time horizons, {\small{POMATO}}'s model core includes additional functionality to identify these redundant constraints.
Implementation and performance of this redundancy-removal algorithm has been presented in our previous work \cite{weinhold2020fast}, where we showed that for most of the studied transmission networks over 99\% of the constraints can be removed from the optimization problem, allowing to solve previously infeasible larger market models and significantly reducing the solve time of smaller problems. 

The methods and algorithms implemented in the model core require a suitable framework that enables fast robust computations and facilitates the use of interchangeable solver back-ends.
The open-source Julia Language provides a competitive combination of performance and readability, \cite{bezanson2017julia}, as well as suitable libraries for implementing the required mathematical programming (electricity market model and redundancy removal). Here, we opted for the popular, well-readable and flexible JuMP package, \cite{DunningHuchetteLubin2017}.
{\small{POMATO}}'s Julia modules (\texttt{MarketModel.jl} and \texttt{RedundancyRemoval.jl}) are parametrized and called automatically by the higher {\small{POMATO}} layers, as indicated in Figure~\ref{fig:flowchart}.

\begin{remark}[Chance Constrained Economic Dispatch]
When modeling systems with high shares of intermittent renewable generation, internalisation of the feed-in uncertainty with respect to capacity reserve and network constraints can be done using Chance Constraints, \cite{dvorkin2019chance,mieth2020risk}. This capability is also included in {\small{POMATO}} as an experimental feature and subject to future publication. 
\end{remark}

\begin{remark}[Solvers]
JuMP allows to easily communicate with a wide range of solvers. 
Per default, {\small{POMATO}} will install and use the open CLP solver\footnote{\url{github.com/jump-dev/Clp.jl}}.   
However, if available on the user's system, {\small{POMATO}} can be configured to use commercial solvers like Mosek\footnote{\url{github.com/MOSEK/Mosek.jl}} or Gurobi\footnote{\url{github.com/jump-dev/Gurobi.jl}}.
\end{remark}

\subsection{Data Management and Pre-Processing}\label{ssec:data_layer}

{\small{POMATO}}'s central Data-Management layer, see Figure~\ref{fig:flowchart}, manages the functionality of the model core by providing data and model parametrizations, calling and monitoring the Julia processes, and validating their results. 
To enable flexible data-handling that avoids redundant calculations, we leverage Python and its object-oriented programming paradigm which resolves issues typically related to static scripts. 
In such linear approaches, e.g. in GAMS, all parameters are handled in the global scope of the program so that any desired conditional functionality renders the code overly complex and hard to maintain. Alternatively, encapsulating the required functionally in different objects (modules), ensures data consistency and compatible methods.

The data-management layer consists of four main modules, \texttt{DataManagement}, \texttt{GridTopology}, \texttt{GridModel} and \texttt{MarketModel}. 
These modules are created once in every instance of {\small{POMATO}} and each provides specific methods to the user-interface and other modules. 
Additionally, single-use modules are used for result processing, e.g. FBMC calculations and result visualization.
\texttt{DataManagement} is the central module to handle all data and their corresponding options.
All input data is validated with respect to predefined data structures that are required to run the desired market model and are made available to other modules. 
Due to the persistence of \texttt{DataManagement}, changes in data, e.g. from user manipulations or model results, are propagated throughout {\small{POMATO}}. 

Pre-processing and validation of data related to the transmission network topology requires specific methods that are collected in the \texttt{GridTopology} module. It uses nodes and lines data and provides the parameters for subsequent power flow calculations. 
Additionally, \texttt{GridTopology} verifies the properties of the network graph, sets reference nodes for multiple disconnected networks, creates contingency scenarios and manages settings of phase shifting transformers.

The verified and pre-processed data is then used by the \texttt{GridModel} and \texttt{MarketModel} modules to provide the input data for the model core. 
Based on the market model that is desired from the user (and that is feasible given the available data), \texttt{GridModel} uses the data provided from \texttt{GridTopology} and \texttt{DataManagement} to compile the correct grid representation with all physical properties.
This includes the parametrization of the network constraints, as discussed in Section~\ref{ssec:model_core} above, as well as generator locations, cost models, and zonal configurations.
Also, \texttt{GridModel} calls \texttt{RedundancyRemoval} in the process of compiling the set of parameters for a N-1 secure dispatch. If this algorithm has been performed in previous runs, \texttt{GridModel} loads available parameters to avoid time-consuming re-runs. 
These parameters define the constraints under which the market has to be cleared. Finally, \texttt{MarketModel} collects these constraints and manages the required model runs. 
After the optimization in the model core has obtained an optimal solution, \texttt{MarketModel} instantiates a new \texttt{Results} object in the \texttt{DataManagement} module.
Communication between the {\small{POMATO}} instance and the mathematical core is achieved via a threaded process within the \texttt{JuliaDaemon} module, see Remark~\ref{rem:julia_daemon}.

The decoupling of model runs and data processing allows to efficiently solve multi-stage market clearing processes that iterate between different market model configurations. Two relevant examples are itemized below.\\

\begin{figure}[ht]
	\centering
    \includegraphics[width=0.8\textwidth]{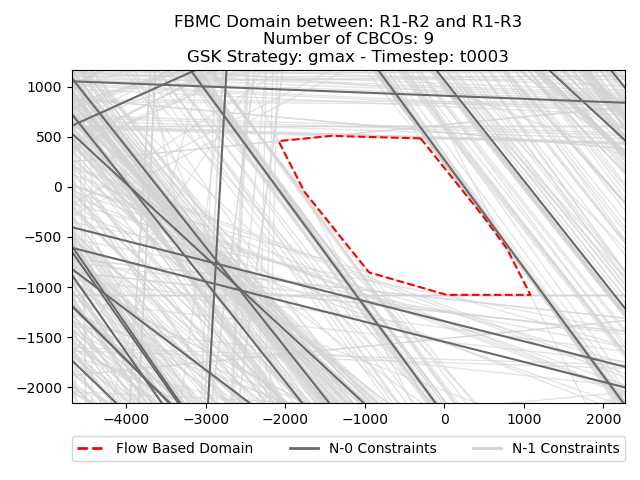}
	\caption{Exemplary Flow Based Domain}
	\label{fig:fbmc_domain}
\end{figure}

\begin{minipage}{0.6\textwidth}
    \begin{enumerate}[Ex. 1 -]
        \item 
        \textit{Redispatch Analysis:~}
        Zonal electricity markets are cleared as a single price zone without any internal network constraints. In a second stage, potential transmission line overloads are rectified by out-of-market CM measures, i.e. redispatch. For this purpose, a second model run with nodal resolution is required. 
        \\
        \\
        \item
        \textit{FBMC:~}
        FBMC uses a forecasted nodal dispatch (base case) to derive zonal PTDF matrices that establish the relation between zonal net-positions and flows on critical network elements (CBCOs, critical branches under critical outages). 
        The resulting feasible region of net-positions with respect to the technical limits of the CBCOs are called \textit{FBMC domains}.
        These domains are used to solve the zonal market clearing that is then corrected by subsequent CM measures.
        To automatize the necessary re-parametrization of parts of the model data, including the generation of the \textit{FBMC domains} depicted in Figure \ref{fig:fbmc_domain}, and re-runs of the optimization model, the additional \texttt{FBMCModule} has been implemented.
    \end{enumerate}
\end{minipage}
\begin{minipage}[t]{0.05\textwidth}
\hspace{\fill}
\end{minipage}
\begin{minipage}{0.25\textwidth}
    \includegraphics[width=1.2\textwidth]{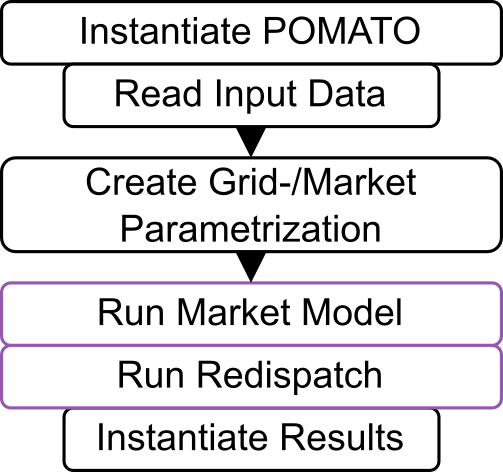}\\
    \vspace{.9cm}
    \includegraphics[width=1.2\textwidth]{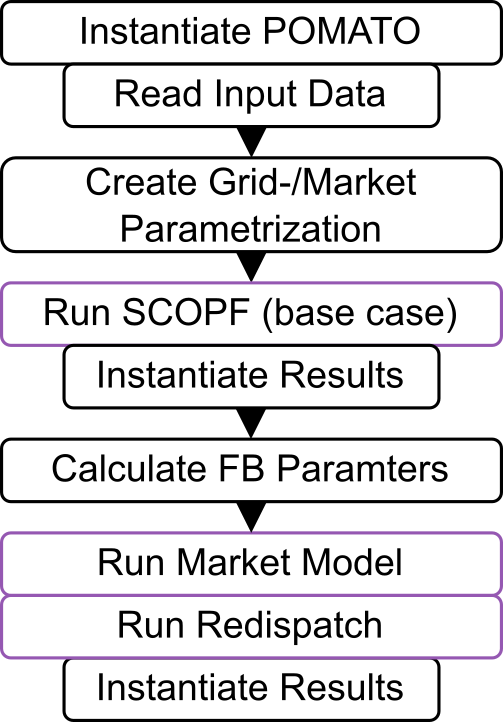}\\
    \vspace{2.4cm}
\end{minipage}

\begin{remark}[Contingency Scenarios]
Alternatively to a strict N-1 contingency analyses, i.e. protecting the system against the outage of every single transmission line, the user can specify customized contingency scenarios.
Such scenarios can either ignore certain lines as potential contingencies or define contingency groups of outages that are likely to occur simultaneously, e.g. parallel lines. 
\end{remark}

\begin{remark}[Python-Julia Interface]\label{rem:julia_daemon}
Communication between the Julia language and Python is implemented by a \texttt{JuliaDaemon} class that maintains a threaded Julia program and executes its functions when requested by \texttt{MarketModel} or \texttt{GridModel}.
Instructions are transferred via \textit{json} files that are written to the {\small{POMATO}} package folder.
This way, the Julia process has to be started only once and remains alive after a model is completed, which allows fast re-solving of a model with different parameters.
\end{remark}

\subsection{User Interface}

All modules of the data processing layer presented in previous Section~\ref{ssec:data_layer} are attributed to the overall {\small{POMATO}} object, which provides a comprehensive set of functions to the user.
Although the methods of the lower level modules are accessible to the user, the {\small{POMATO}} module provides predefined methods for the typical use cases of this program. 
A brief overview is presented in the illustrative application in Section~\ref{sec:application}.
Additionally, {\small{POMATO}} provides a collection of visualization features. 
The \texttt{GeoPlot} module shows market and redispatch results including prices on an interactive map using the \textit{Bokeh}\footnote{\url{bokeh.org}} library.
Additionally, many results, such as the FBMC domains, can be automatically plotted as a collection of 2D diagrams. 
\subsection{Publication}

{\small{POMATO}} is available under the \textit{LGPLv3} license on GitHub including a comprehensive documentation, \cite{pomato_repo,pomato_doc}.
The repository also includes exemplary data-sets that cover the core functionality. 
These include a data-set for the German power system, that has been compiled to showcase {\small{POMATO}}'s functionality, as well as the NREL 118 bus network, \cite{nrel_118_2018}, and Matpower case files. 
More details on how to create custom data-sets can be found in the accompanying documentation\footnote{\url{pomato.readthedocs.io/en/latest/input_model_data.html}}.

\begin{remark}(Installation)
{\small{POMATO}} requires Python and Julia to be installed.
Installation of the Python modules is managed via pip\footnote{\url{pip.pypa.io}}.
If a working Julia distribution is detected on the system, the two Julia packages, MarketModel and RedundancyRemoval are automatically pulled from their separate repositories and installed during this process.
\end{remark}

\section{Illustrative Application}\label{sec:application}
This section illustrates the application of {\small{POMATO}} and the corresponding example code is shown below. 

\begin{minted}[bgcolor=bg,linenos]{python}
    from pomato import POMATO
    
    mato = POMATO(wdir=wdir, options_file="options/de.json")
    mato.load_data('data_input/dataset_de.zip')
    
    mato.create_grid_representation()
    mato.run_market_model()
    
    market_result, redispatch_result = mato.data.return_results()
    
    # Check for Overloaded lines N-0
    n0_m, _ = market_result.overloaded_lines_n_0()
    print("Number of N-0 Overloads: ", len(n0_m))
    n0_r, _  = redispatch_result.overloaded_lines_n_0()
    print("Number of N-0 Overloads: ", len(n0_r))
    n1_r, _  = redispatch_result.overloaded_lines_n_1()
    print("Number of N-1 Overloads: ", len(n1_r))
    
    gen = redispatch_result.redispatch()
    print("Total Redispatch in MWh: ", gen.delta_abs.sum())

    mato.create_geo_plot(title="DE: Redispatch")
\end{minted}

\begin{figure}[ht]
	\centering
    \includegraphics[width=\textwidth]{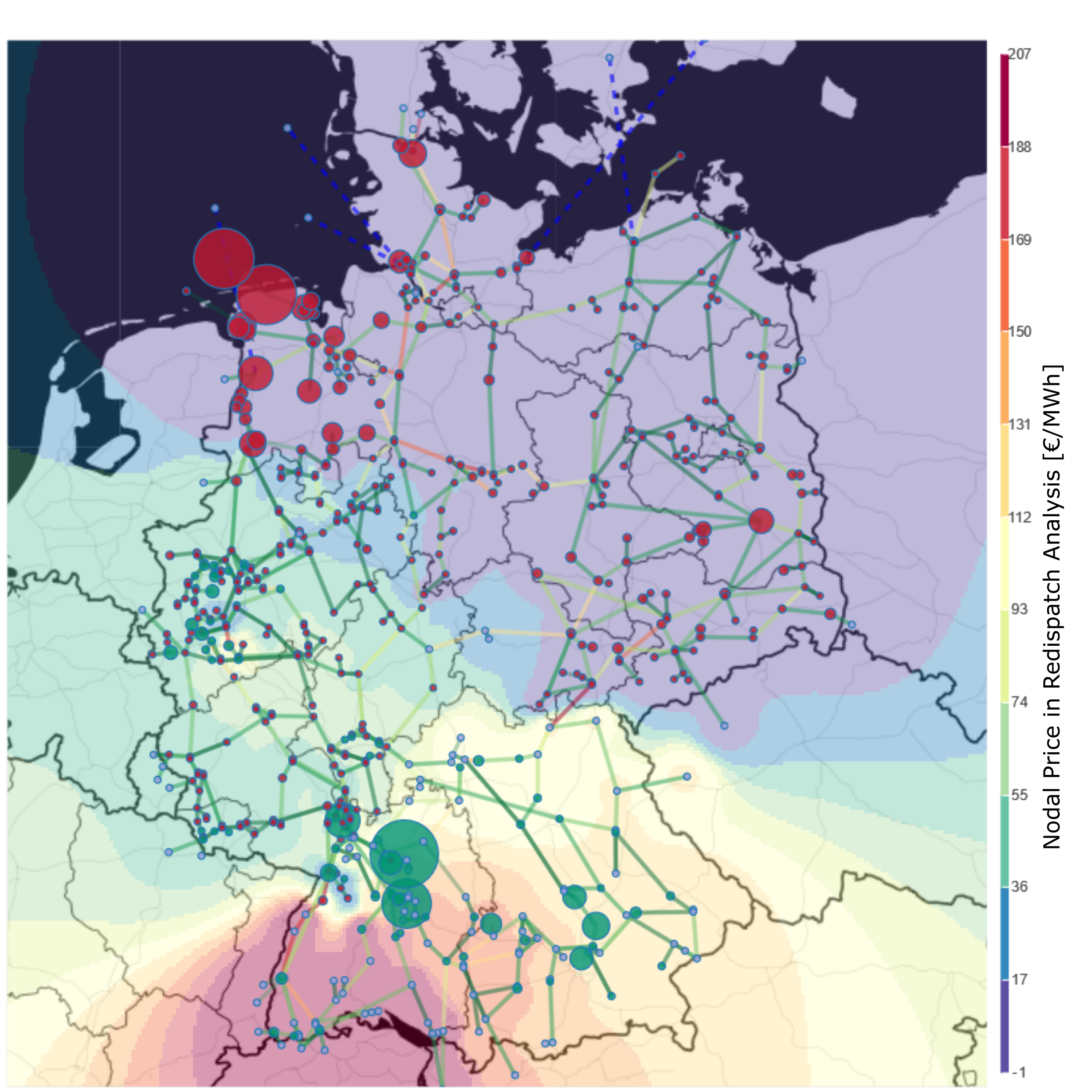}
	\caption{GeoPlot of market- and redispatch result for Germany. Red and green circles indicate negative and positive changes in generation schedules, respectively, due to redispatch. Change magnitude are proportional to the circles' radius. Line colors indicate average line load where green is low loads and red is high load. Prices as dual multipliers on the nodal power balances of the redispatch run are indicated by the contour plot overlay.}
	\label{fig:geoplot_de}
\end{figure}

After importing the {\small{POMATO}} module (\texttt{1}), a new {\small{POMATO}} class is instantiated in \texttt{3} with the requested working directory (\texttt{wdir}) and an options file in the JSON\footnote{\url{www.json.org}} format that configures the model run.
In this example the  options file specifies a zonal market clearing and subsequent redispatch with redispatch-cost and model horizon:
\begin{minted}[bgcolor=bg]{json}
	{"optimization": {
		"type": "dispatch",
		"model_horizon": [0, 168],
		"redispatch": {
			"include": true,
			"cost": 50}}
\end{minted}
A comprehensive list and descriptions of possible configurations can be found in the online documentation\footnote{\url{pomato.readthedocs.io/en/latest/options.html}}. 
In \texttt{4} the {\small{POMATO}} object loads the required data, which automatically instantiates the data management objects as described in Section~\ref{ssec:data_layer} above.
The command in \texttt{6} creates the grid representation for the specified model run and \texttt{7} starts the actual model run. 
After successful completion, two \texttt{Results} are instantiated into \texttt{DataManagement} and are available to the user (\texttt{9})  for further analysis, e.g. to check for overloaded lines with and without contingencies (\texttt{11}--\texttt{20}). 
Finally the \texttt{GeoPlot} is created in \texttt{22} which yields Figure \ref{fig:geoplot_de}.

Execution time of this example is about 7.5 minutes, including data read-in, Julia start-up, solving both the market- and the redispatch problem, as well as the final result analysis and plotting. Computation times of subsequent runs will be shorter due to already active Julia process.

\section{Conclusion and further development}

This paper presents a tool that aims to enable the broad analyses of zonal electricity markets. The proposed Power Market Tool {\small{POMATO}} is open-source, provides essential data-sets and comprehensive online documentation.
Central to {\small{POMATO}}'s functionality is its ability to synthesize the flow-based market-coupling (FBMC) process. Further {\small{POMATO}} includes a redundancy removal algorithm that enables security constrained (N-k) dispatch with feasible computation times. 
While the current version of {\small{POMATO}} already includes prototypical options for the analyses of stochastic (chance-constrained) market clearing, extensions to stochastic and risk-aware markets and power flow analyses along the lines of \cite{dvorkin2019chance,mieth2020risk} are subject to ongoing development. 

\FloatBarrier
\section*{Conflict of Interest}
We wish to confirm that there are no known conflicts of interest associated with this publication and there has been no significant financial support for this work that could have influenced its outcome.

\section*{Acknowledgements}
We gratefully acknowledge the support by the German Federal Ministry for Economic Affairs and Energy (BMWi) in the project “Long-term Planning and Short-term Optimization of the German Electricity System Within the European Context” (LKD-EU, 03ET4028A). The work of R. Weinhold was supported by the Danish Energy Agency. The work of R. Mieth was supported by the Reiner Lemoine-Foundation.

\clearpage
\bibliographystyle{elsarticle-num}
\bibliography{bibliography.bib}

\begin{thebibliography}{10}
\expandafter\ifx\csname url\endcsname\relax
  \def\url#1{\texttt{#1}}\fi
\expandafter\ifx\csname urlprefix\endcsname\relax\def\urlprefix{URL }\fi
\expandafter\ifx\csname href\endcsname\relax
  \def\href#1#2{#2} \def\path#1{#1}\fi

\bibitem{amprion_fbmc}
Amprion, {Flow Based Market Coupling -- Development of the Market and Grid
  Situation 2015-2017},
  \newline\url{amprion.net/Dokumente/Dialog/Downloads/Studien/CWE/CWE-Studie_englisch.pdf}
  (2018).

\bibitem{eu_2015_1222}
{The European Commission}, {Commission Regulation ({EU}) no 2015/1222},
  \newline\url{eur-lex.europa.eu/legal-content/EN/TXT/HTML/?uri=CELEX:32015R1222\&from=EN}
  (2015).

\bibitem{amprion2011}
Amprion, APX-ENDEX, Belpex, Creos, Elia, EnBW, {EPEX SPOT}, RTE, TenneT, {CWE
  Enhanced Flow-Based MC feasibility report},
  \newline\url{www.amprion.net/Dokumente/Strommarkt/Engpassmanagement/CWE-Market-Coupling/CWE_FB-MC_feasibility_report.pdf}
  (March 2011).

\bibitem{ec2016/1719}
E.~Commission, Commission regulation (eu) 2016/1719 of 26 september 2016
  establishing a guideline on forward capacity allocation, Official Journal of
  the European Union~(September 26) (2016).

\bibitem{ehrenmann2005inefficiencies}
A.~Ehrenmann, Y.~Smeers, Inefficiencies in european congestion management
  proposals, Utilities policy 13~(2) (2005) 135--152.

\bibitem{schonheit2020impact}
D.~Sch{\"o}nheit, R.~Weinhold, C.~Dierstein, The impact of different strategies
  for generation shift keys (gsks) on the flow-based market coupling domain: A
  model-based analysis of central western europe, Applied Energy 258 (2020)
  114067.

\bibitem{entsoe_biddingzones}
ENTSO-E, {First Edition of the Bidding Zone Review -- Final Report},
  \newline\url{eepublicdownloads.azureedge.net/clean-documents/news/bz-review/2018-03_First_Edition_of_the_Bidding_Zone_Review.pdf}
  (2018).

\bibitem{fur2018netzentwicklungsplanung}
{Genehmigung des Szenariorahmens f{\"u}r die Netzentwicklungsplanung},
  Netzentwicklungsplanung 2019-2030, Bonn: Bundesnetzagentur f{\"u}r
  Elektrizit{\"a}t, Gas, Telekommunikation, Post und Eisenbahnen (BNetzA)
  (2018).

\bibitem{aravena2016renewable}
I.~Aravena, A.~Papavasiliou, Renewable energy integration in zonal markets,
  IEEE Transactions on Power Systems 32~(2) (2016) 1334--1349.

\bibitem{matthes2019impact}
B.~Matthes, C.~Spieker, D.~Klein, C.~Rehtanz, Impact of a minimum remaining
  available margin adjustment in flow-based market coupling, in: 2019 IEEE
  Milan PowerTech, IEEE, 2019, pp. 1--6.

\bibitem{schonheit2020elmod}
D.~Sch{\"o}nheit, D.~Hladik, H.~Hobbie, D.~M{\"o}st, Elmod documentation:
  Modeling of flow-based market coupling and congestion management, Tech. rep.,
  Working paper of the Chair of Energy Economics (TU Dresden) (2020).

\bibitem{spieker2016european}
C.~Spieker, D.~Klein, V.~Liebenau, J.~Teuwsen, C.~Rehtanz, European electricity
  market and network simulation for energy system analysis, in: 2016 IEEE
  International Energy Conference (ENERGYCON), IEEE, 2016, pp. 1--6.

\bibitem{leuthold2012large}
F.~U. Leuthold, H.~Weigt, C.~von Hirschhausen, A large-scale spatial
  optimization model of the european electricity market, Networks and spatial
  economics 12~(1) (2012) 75--107.

\bibitem{schonheit2020minimum}
D.~Schönheit, C.~Dierstein, D.~Möst, Do minimum trading capacities for the
  cross-zonal exchange of electricity lead to welfare losses?, Energy Policy
  (2020).

\bibitem{DunningHuchetteLubin2017}
I.~Dunning, J.~Huchette, M.~Lubin, Jump: A modeling language for mathematical
  optimization, SIAM Review 59~(2) (2017) 295--320.

\bibitem{pomato_repo}
{Pomato - GitHub Repository}, \newline\url{github.com/richard-weinhold/pomato}
  (2020).

\bibitem{pomato_doc}
{Pomato - Documentation}, \newline\url{pomato.readthedocs.io/en/latest} (2020).

\bibitem{weinhold2020fast}
R.~Weinhold, R.~Mieth, Fast security-constrained optimal power flow through
  low-impact and redundancy screening, IEEE Transactions on Power Systems
  35~(6) (2020) 4574--4584.

\bibitem{horsch2018}
J.~H{\"o}rsch, H.~Ronellenfitsch, D.~Witthaut, T.~Brown, Linear optimal power
  flow using cycle flows, Electric Power Systems Research 158 (2018) 126--135.

\bibitem{bouffard2005umbrella}
F.~Bouffard, F.~D. Galiana, J.~M. Arroyo, Umbrella contingencies in
  security-constrained optimal power flow, in: 15th Power systems computation
  conference, PSCC, Vol.~5, 2005.

\bibitem{bezanson2017julia}
J.~Bezanson, A.~Edelman, S.~Karpinski, V.~B. Shah, Julia: A fresh approach to
  numerical computing, SIAM review 59~(1) (2017) 65--98.

\bibitem{dvorkin2019chance}
Y.~Dvorkin, A chance-constrained stochastic electricity market, IEEE
  Transactions on Power Systems (2019).

\bibitem{mieth2020risk}
R.~Mieth, J.~Kim, Y.~Dvorkin, Risk-and variance-aware electricity pricing,
  Electric Power Systems Research 189 (2020) 106804.

\bibitem{nrel_118_2018}
I.~Pena, C.~B. Martinez-Anido, B.-M. Hodge, An extended ieee 118-bus test
  system with high renewable penetration, IEEE Transactions on Power Systems
  33~(1) (2017) 281--289.

\end{thebibliography}

\end{document}